\documentclass{article} 
\usepackage{amsfonts}
\usepackage{amsmath}
\usepackage{amssymb} 
\newcommand{\nc}{\newcommand}
\allowdisplaybreaks[2] \nc{\fL}{\frak{L}}
 
\renewcommand{\Im}{\mathrm{Im}\,} \nc{\di}{\displaystyle}
\nc{\nn}{\nonumber} \nc{\nek}{\nonumber\\[1ex]} \nc{\st}{\scriptstyle}
\nc{\tst}{\textstyle} \nc{\nb}{\normalsize\bf} \nc{\ns}{\normalsize}
\nc{\seq}{\subseteq} \nc{\AH}{\mathcal{A}} \nc{\BE}{\mathcal{B}}
\nc{\CE}{\mathcal{C}} \nc{\D}{\mathcal{D}} \nc{\E}{\mathcal{E}}
\nc{\EF}{\mathcal{F}} \nc{\GE}{\mathcal{G}} \nc{\HA}{\mathcal{H}}
\nc{\J}{\mathcal{J}} \nc{\KA}{\mathcal{K}} \nc{\EL}{\mathcal{L}}
\nc{\PE}{\mathcal{P}} \nc{\ER}{\mathcal{R}} \nc{\ES}{\mathcal{S}}
\nc{\TE}{\mathcal{T}} \nc{\EM}{\mathcal{M}} \nc{\EN}{\mathcal{N}}
\nc{\OH}{\mathcal{O}} \nc{\U}{\mathcal{U}} \nc{\WE}{\mathcal{W}}
\nc{\EX}{\mathcal{X}} \nc{\Y}{\mathcal{Y}} \nc{\ZE}{\mathcal{Z}}
\nc{\ma}[1]{\mbox{$\,{#1}\,$}} \nc{\ek}{\protect\\[1ex]}
\nc{\zx}{\protect\\[2ex]} \newcommand{\C}{{\mathbb C}}

 \newcommand{\R}{{\mathbb R}}
 
 \nc{\IR}{\mbox{\bf R}} \nc{\IN}{\mbox{\bf N}}
\nc{\ZZ}{\mbox{\bf Z}} \nc{\la}{\lambda} \nc{\La}{\Lambda}
\nc{\da}{\delta} \nc{\Da}{\Delta} \nc{\ta}{\theta} \nc{\Ta}{\Theta}
\nc{\na}{\nabla} \nc{\ue}{\infty} \nc{\vp}{\varphi} \nc{\vta}{\vartheta}
\nc{\Gm}{\Gamma} \nc{\gm}{\gamma} \nc{\ka}{\kappa} \nc{\si}{\sigma}
\nc{\Si}{\Sigma} \nc{\al}{\alpha} \nc{\be}{\beta} \nc{\om}{\omega}
\nc{\Om}{\Omega} \nc{\pa}{\partial} \nc{\ti}{\times} \nc{\n}{|}
\nc{\rub}{\,\rule[-2.7pt]{.02in}{4mm}\,} \nc{\ab}{\|} \nc{\s}{\tilde}
\nc{\ve}{\varepsilon} \nc{\fa}{\forall} \nc{\ov}{\overline}
\nc{\un}{\underline} \nc{\Llr}{\Longleftrightarrow}
\nc{\llr}{\longleftrightarrow} \nc{\ra}{\rightarrow}
\nc{\lra}{\longrightarrow} \nc{\rh}{\rightharpoonup} \nc{\Ra}{\Rightarrow}
\nc{\ran}{\rangle} \nc{\lan}{\langle} \nc{\bs}{\backslash} \nc{\ko}{\,,\,}
\nc{\eq}[1]{\mbox{\rm {(\ref{E#1})}}} \newcommand{\qed}{\mbox{
}\nolinebreak\hfill \rule{2mm}{2mm}} \nc{\ha}{\frac{1}{2}}
\nc{\lk}{\left[} \nc{\rk}{\right]} \nc{\lb}{\left\{} \nc{\rb}{\right\}}
\nc{\rr}{\right)} \nc{\lr}{\left(} \nc{\f}{\big(} \nc{\g}{\big)}
\nc{\Ba}{\Big(} \nc{\Bz}{\Big)} \nc{\Bka}{\Big[} \nc{\Bkz}{\Big]}
\nc{\bka}{\big[} \nc{\bkz}{\big]} \nc{\Blb}{\Big\{} \nc{\Brb}{\Big\}}
\nc{\blb}{\big\{} \nc{\brb}{\big\}} \nc{\pn}{\par\noindent}
\nc{\emp}{\emptyset} \nc{\Ri}{\Rightarrow}
\nc{\rain}[1]{\raisebox{-3pt}{${\tst |}_{#1}$}} \nc{\hph}{\hphantom}
\nc{\vph}{\vphantom} \nc{\vpn}{\vspace{2ex}\par\noindent}
\nc{\vpar}{\vspace{2ex}\par} \nc{\mathe}[1]{\mbox{${\di {#1}}$}}
\nc{\mathr}[2]{{\di\mathrel{\mathop{#1}_{#2}}}} \nc{\sD}{{\tilde{D}}}
\newtheorem{lemma}{Lemma}
\newtheorem{theorem}{Theorem}
\newtheorem{corollary}{Corollary}

\numberwithin{equation}{section} 
\begin{document} 
 \markboth{A. G. Ramm }{Some results on inverse scattering}
 
\title{SOME RESULTS ON INVERSE SCATTERING} \author{A. G. RAMM
\footnote{Mathematics Department, Kansas State University, 
Manhattan, KS 66506, USA}} 


\maketitle %


\begin{abstract} 
A review of some of the author's results in the area of inverse
scattering is given. The following topics are discussed: 1) Property $C$
and applications, 2) Stable inversion of fixed-energy 3D scattering data
and its error estimate, 3) Inverse scattering with ''incomplete`` data,
4) Inverse scattering for inhomogeneous Schr\"odinger equation, 5) Krein's
inverse scattering method, 6) Invertibility of the steps in
Gel'fand-Levitan, Marchenko, and Krein inversion methods, 7) The 
Newton-Sabatier and Cox-Thompson procedures are not inversion
methods, 8) Resonances: existence,
location, perturbation theory, 9) Born inversion as an ill-posed problem,
10) Inverse obstacle scattering with fixed-frequency data, 11) Inverse
scattering with data at a fixed energy and a fixed incident direction, 
12) Creating materials with a desired refraction coefficient and 
wave-focusing properties. 
\end{abstract} 
\pn
\section{Introduction}\label{SI} This paper
contains a brief description of some of the author's results in the
 area of inverse scattering. The proofs are omitted if the results were 
published. In this case references are given. For some new results proofs 
are given. The following  problems are discussed:

1) Property $C$ and applications, 

2) Stable inversion of fixed-energy 3D scattering data
and its error estimate, 

3) Inverse scattering with ''incomplete`` data,

4) Inverse scattering for inhomogeneous Schr\"odinger equation, 

5) Krein's inverse scattering method, 

6) Invertibility of the steps in
Gel'fand-Levitan, Marchenko, and Krein inversion methods, 

7) The Newton-Sabatier and Cox-Thompson procedures are not inversion 
methods, 

8) Resonances: existence,
location, perturbation theory, 

9) Born inversion as an ill-posed problem,

10) Inverse obstacle scattering with fixed-frequency data, 

11) Inverse scattering with data at a fixed energy and a fixed incident 
direction,

12) Creating materials with a desired refraction coefficient and
wave-focusing properties.

\section{Property $C$ and applications}\label{S1}
Property $C$, completeness of the set of products of solutions to
homogeneous equations, was introduced first in \cite{R196} and then
applied to many inverse problems: 3D inverse scattering with
fixed-energy data, inverse boundary-value problem, inverse problems for
the heat and wave equations, impedance tomography problem, etc. (see
\cite{R196}$^-$\cite{R228},
\cite{R278}, \cite{R470} and references therein).

{\bf Definition 1.} {\it If $L_1$ and $L_2$ are two linear partial 
differential expressions 
(PDE),
$D\subset \R^n$, $n\geq 2$, $N_j = \{ u:L_ju = 0\mbox{ in } D\}$,
$j=1,2$, then the pair $\{ L_1,L_2\}$ has Property $C$ if the set $\{
u_1,u_2\}_{\forall u_j\in N_j}$ is total in $L^2(D)$.} 

Necessary and
sufficient condition for a pair $\{ L_1,L_2\}$ of PDE with constant
coefficients to have Property C is given in \cite{R262}, (see also 
\cite{R278}, \cite{R470}). This condition is easy to verify.
If $\mathcal{L_j}:=\{z: z\in \C^n,\, L_j(z)=0\}$ is the algebraic variety, 
corresponding to the PDE
$L_j$ with constant coefficients and characteristic polynomial $L_j(z)$,
then the necessary and sufficient condition for a pair $\{ L_1,L_2\}$ of 
PDE with constant coefficients to have Property C can be stated as 
follows: the union of the algebraic varieties $\mathcal{L_1}$
and  $\mathcal{L_2}$ is not a union of parallel hyperplanes in 
$\C^n$.  
If the pair $\{ L,L\}$ has property $C$, then 
we say that
the operator $L$ has this property. Classical operators $\na^2$, $\pa_t
- \na^2$, $\pa^2_{tt}-\na^2$, $i\pa_t-\na^2$ all have property $C$,
Schr\"odinger pair $\{ L_1,L_2\}$ has property $C$, where $L_j = \na^2 +
k^2 - q_j(x)$, $k = {\rm const} \geq 0$, $j = 1,2$, $q_j\in Q_a:= \bigl\{
q_j(x)\in L^2(B_a)$, $q_j =\ov{q}_j$, $B_0:= \{ x:\, \n x\n \leq a$,
$x\in \R^2\}$, $q_j = 0$ if $\n x\n > a\bigr\}$ (\cite{R278}, \cite{R470}).

{\bf Example of application of property $C$ to inverse scattering.} Let
$A_q (\al',\al,k):= A(\al',\al)$, $k = {\rm const} > 0$ is fixed, be the
scattering amplitude corresponding to $q\in Q_a$. The inverse scattering
problem with 3D fixed energy data consists of finding $q$ given
$A(\al',\al)$ for all $\al'$, $\al\in S^2$, $S^2$ is the unit sphere in
$\R^3$. This problem has been open for several decades (from 1942). 
Below we write $\be$ in place of $\al'$ sometimes.
In 1987 the author proved that $q\in Q_a$ is uniquely determined by
$A(\al',\al)$ (\cite{R214}, \cite{R228}). The idea of the proof is
simple. One uses the formula (see \cite{R278}, p. 67): 
$$-4\pi A(\be,\al) =
\int_{\R^3}p(x)u_1(x,\al)u_2(x,\be)dx,$$ 
where $A:= A_1-A_2$, $p(x):=
q_1-q_2$, $u_j$ is the scattering solution, corresponding to $q_j$, $j =
1,2$, and $A_j$ is the corresponding scattering amplitude. This formula
is derived by using the formula 
$$G(x,y) = \frac{e^{ik\n y\n}}{4\pi\n
y\n}\, u(x,\al) + o\f \frac{1}{\n y\n}\g, \quad \n y\n \ra \ue,\quad
\frac{y}{\n y\n} = -\al,$$ 
which was proved in \cite{R3}. 

If $A_1=A_2$ $\fa \al,\be\in
S^2$ and fixed $k > 0$, then 
$$\int_{B_a} p(x) u_1(x,\al)u_2(x,\be)dx=0\quad \forall \al,\be\in S^2.$$ 
This 
is an orthogonality relation: $p(x)$ is orthogonal to the set 
$\{ u_1 u_2\}_{ \forall \al,\be \in S^2}$. The set $\{ 
u_j(x,\al)\}_{\forall \al\in 
S^2}$ is total in the set $N_j:=\{u:l_j u:= [\na^2+k^2-q_j(x)]u=0$ in 
$B_a\}$. The pair $\{L_1,L_2\}$ has property $C$. Thus, $p(x)=0$, and the 
uniqueness theorem for inverse scattering with fixed-energy data is proved. 

Let us give an example of applications of property $C$ to inverse 
boundary-value problem. 
Let 
$$Lu:=[\na^2 + k^2-q(x)]u=0 \hbox{\,\, in \,\,} D\subset \R^3,$$
where $D$ is a bounded 
domain with a smooth boundary $S$, $u\Big|_S=f$. Assume that 
zero 
is not a Dirichlet eigenvalue of $L$ in $D$. Then $f$ defines $u=u(x; f)$ 
uniquely, and the map $\Lambda :f\to h:=u_N$, where $u_N$ is the normal 
derivative of $u$ on $S$, is well defined. The inverse problem is: 

{\it Given the set $\{f,h\}_{\forall f\in H^{3/2}(S)}$ can one determine 
$q$ uniquely?} 

The answer is yes. Indeed, if $q_1$ and $q_2$ generate the same 
set $\{f,h\}_{\forall f\in H^{3/2}(S)}$, i.e., the same $\Lambda$, then one 
derives, as above, the orthogonality relation 
$$\int_D p(x)u(x;f)v(x)dx=0 \quad 
\forall f \in H^{3/2}(S), \,\,\,\forall v \in N_2,$$ 
where $p:=q_1-q_2$. The 
set 
$\{u v\}$ is total in $L^2(D)$ by property $C$ for 
the pair $\{ L_1,L_2\}$. Thus, 
$p=0$, and the uniqueness is proved. Many other examples one finds in \cite{R278}, \cite{R470}.

\section{Stable solution of $3D$ inverse scattering problem}\label{S2}
If $q \in Q:=Q_a\cap L^\ue(B_a)$ and $A(\be,\al)$ is the exact 
scattering 
amplitude at a fixed $k>0$, then it is proved in \cite{R190} that $A$ 
admits an analytic continuation from $S^2 \times S^2$ to the algebraic 
variety $M_k :=\{ \Theta:\Theta\in \C^3, \Theta\cdot\Theta=k^2\}$,
 where $\Theta\cdot\Theta=\sum_{j=1}^3 \Theta_j^2$. This implies that the 
knowledge of $A(\be,\al)$ on $S_1^2 \times S_2^2$, where $S_j^2, j=1,2$, are 
arbitrary small open subsets of $S^2$, determines $A(\be,\al)$
uniquely on $S^2\times S^2$.

For any $\xi\in \R^3$ there exist (many) $\Theta', \Theta\in M_k$ such 
that $\Theta'-\Theta=\xi$, $|\Theta|\to\ue$. We take $k=1$ in this Section 
without loss of generality. Let $M:=M_1$, $\tilde{q}(\xi):=\int_{B_a}
e^{-i\xi\cdot x} q(x)dx$, $Y_\ell(\Theta)=Y_{\ell m}(\Theta)$, 
$-\ell \le m \le\ell$, are orthonormal spherical 
harmonics, $A_\ell(\al):=\int_{S^2} A(\be,\al)\overline{Y_\ell(\be)}d\be$, 
the 
overbar stands for complex conjugate, 
$h_\ell(r)=e^{i\frac{\pi}{2}(\ell+1)} 
\sqrt{\frac{\pi}{2r}}H_{\ell+\ha}^{(1)}(r)$, $H_\ell^1(r)$ is the 
Hankel function, $\sum_{\ell=0}^\ue := \sum_{\ell=0}^\ue 
\sum_{m=-\ell}^\ell$, $b$ and $a_1$ are arbitrary numbers satisfying the 
inequality $b>a_1>a$. Let $\| \rho\| := \| \rho\|_{L^2(B_b\backslash B_{a_1})},
 \rho := e^{-i\Theta\cdot x} \int_{S^2} u(x,\al)\nu(\al)d\al-1$, $\nu\in L^2(S^2)$, $\Theta\in M$. Let $d(\Theta)=\inf_{\nu\in L^2(S^2)} \|\rho(\nu)\|$. One has 
$$d(\Theta)\le c|\Theta|^{-1}, \quad \Theta\in M, \,\, |\Theta|\to \ue,$$ 
where $c=\text{const }>0$ is independent of $\Theta$, see \cite{R228}. Let 
$\Theta',\Theta\in M$, 
$\Theta'-\Theta=\xi$, where $\xi\in\R^3$ is an arbitrary vector. If 
$\|\rho(\nu)\|<2d(\Theta)$, $\nu=\nu(\al,\Theta)$, and 
$$\hat{q}:= -4\pi\int_{S^2}A(\Theta',\al)\nu (\al,\Theta)d\al,$$ 
then 
$$|\tilde{q}(\xi)-\hat{q}|\le\frac{c}{|\Theta|}, \quad |\Theta|\to\ue,$$
where $ c>0$ 
stands for various constants. Thus, (\cite{R470}):
	\[
	-4\pi\lim_{|\Theta|\to\ue,\Theta',\Theta\in M\atop \Theta'-\Theta=
\xi,\|\rho(\nu(\al,\Theta))\|<2d(\Theta)} \int_{S^2} A(\Theta',\al)
y\nu(\al,\Theta)d\al=\tilde{q}(\xi).
\]

If $\max_{\be,\al\in S^2} |A_1(\be,\al)-A_2(\be,\al)|<\da$, then 
$$\max_{\xi\in\R^3} |\tilde{q}_1(\xi)-\tilde{q}(\xi)|<c\frac{\ln|\ln \da|}
{|\ln \da|},$$
as was proved in  \cite{R285}. 

Suppose the "noisy data" $A_\da$ are given, 
$\sup_{\be,\al} |A_\da(\be,\al)-A(\be,\al)|<\da$, where $A_\da$ is not 
necessarily a scattering amplitude, $A(\be,\al)$ is the scattering amplitude 
corresponding to $q\in Q$. The author's method for calculating a stable 
estimate of $\tilde{q}(\xi)$ is as follows.

Let $N(\da):=[\frac{|\ln\da|}{\ln|\ln\da|}],[x]$ is the integer closest 
to $x>0$,
\begin{eqnarray*}
\hat{A}_\da(\Theta',\al)&:=&\sum_{\ell=0}^{N(\da)} A_{\da \ell} (\al)
Y_\ell(\Theta'),\,\,\,  u_\da(x,\al):=e^{i\al\cdot 
x}+\sum_{\ell=0}^{N(\da)} A_{\da\ell}(\al)Y_\ell(x^0)h_\ell(r),\\
&& r:=|x|, \, x^0:=\frac{x}{r},\,\,\Theta',\Theta\in M, 
\Theta'-\Theta=\xi\\
\rho_\da&:=&e^{-i\Theta\cdot x}\int_{S^2} 
u_\da(x,\al)\nu(\al)d\al-1, \,\,\nu\in 
L^2(S^2),\\  
\mu(\da)&:=&e^{-\gm N(\da)}, \gm:=\ln\frac{a_1}{a}>0,\,\, 
a(\nu):=\|\nu\|_{L^2(S^2)}, \kappa=|\Im \Theta|,\\
F(\nu,\Theta)&:=&\|\rho_\da(\nu)\| + a(\nu)e^{\kappa b}\mu(\da)=\inf_{\nu\in 
L^2(S^2)\atop \Theta'-\Theta=\xi,\Theta',\Theta\in M,|\Theta|\gg 1}:=t(\da),\\
&&\tau(\da):= \frac{(\ln |\ln \da|)^2}{|\ln\da|}.
\end{eqnarray*}
We prove that $t(\da)=O(\tau(\da))$ as $\da\to 0$, and $t(\da)$ is 
independent of $\xi$. If $\nu=\nu_\da (\al)\in L^2(S^2)$ and 
$\Theta=\Theta_\da$ are such that $F(\nu_\da,\Theta_\da)<2t(\da)$, and 
$$\hat{q}_\da:=-4\pi\int_{S^2}\hat{A}_\da(\Theta'_\da,\al)\nu_\da(\al)d\al,$$ 
then $$\sup_{\xi\in\R^3}|\hat{q}_\da-\tilde{q}(\xi)|\le 
c\tau(\da),\quad 0<\da\ll 1,$$ where $c=\text{const}>0$ depends only on a 
norm 
of $q$ (see \cite{R425} and \cite{R531}). 
\section{Inverse scattering with ''incomplete`` data}\label{S3}
\subsection{Spherically symmetric potentials} Let $q\in Q$ be spherically symmetric. It was proved in
\cite{R241}, \cite{R260} that a necessary and sufficient condition for
$q\in Q$ to be spherically symmetric is $A(\be,\al)=A(\be\cdot \al)$. It
was known for decades that if $q(x)=q(\n x\n)$ then
$A(\be,\al)=A(\be\cdot \al)$. This follows easily from the separation of
variables. The converse is a non-trivial fact, which follows from the
author's uniqueness theorem for inverse scattering problem. If $q=q(r)$,
$r = \n x\n$, and $q\in Q$, then the knowledge of the scattering
amplitude $A(\be\cdot \al)$ is equivalent to the knowledge of all
fixed-energy phase shifts $\da_\ell$, $A_\ell(\al) = A_\ell
Y_\ell(\al)$, $A_\ell = 4\pi e^{i\da_\ell}\, \sin \da_\ell$, $k = 1$.
One can find the radius $a$ of the ball $B_a$, out of which $q = 0$, by 
the
formula (\cite{R470}, p. 173): $$a = 2e^{-1}\lim_{\ell\ra \ue}(\ell \n
\da_\ell\n^{\frac{1}{2\ell}}),$$ and the author has proved that if
$\mathfrak{L}$ is any subset of positive integers such that 
\begin{equation}
\label{E1s}
\sum_{\ell \in \frak{L}}\ell^{-1} = \ue,
\end{equation}
 then the set $\{
\da_\ell\}_{\ell\in \fL}$
 determines $q\in Q$ uniquely (\cite{R393}). He
conjectured that \eq{1s} is necessary for the uniqueness of the recovery
of $q$. This conjecture was proved in \cite{H}. Examples of two quite
different, piecewise-constant potentials $q_j(r)$, generating practically
the same sets of fixed-energy phase shifts, are given in \cite{R421},
\cite{R392}. "Practically the same" means here that 
$\max_{\ell \geq 0}|\da_\ell^{(1)}-\da_\ell^{(2)}|<10^{-5}$.

\subsection{Rapidly decaying potentials}\label{SS3.2} 
To find $q(x)\in L_{1,1}:= \{ q:\, q =
\ov{q},$ $\int_0^\ue x \n q(x)\n dx < \ue\}$ in the 1D inverse
scattering problem on a half-line one needs the following scattering
data
$$\ES:= \{ S(k):= \frac{f(-k)}{f(k)}\Big\n_{\fa k\geq 0},\quad
k_j,s_j,\, 1 \leq j\leq J\},$$ 
where $f(k)$ is the 
Jost function, $k_j>0$ are constants,  $-k_j^2$  are the bound states, 
$f(ik_j) = 0$, $1 \leq j
\leq J$, $s_j > 0$ are the norming constants, 
$$s_j = -\frac{2ik_j}{\dot{f}(ik_j)\, f'(0,ik_j)}\,,$$ 
where 
$$f(x,k) = e^{ikx} +\int_x^\ue A(x,y)e^{iky}dy$$ 
is the Jost solution,  $A(x,y)$ is the
transformation kernel, $f(k):=f(0,k)$, $\dot{f}(k):=\frac {df}{dk}$,
$f'(0, k):=\frac {df(x,k)}{dx}\big|_{x=0}$. 

Given $\ES$, one solves the Marchenko (M) equation
for $A(x,y)$:
\begin{equation}
\label{E3.1}
A(x,y) + \int_x^\ue A(x,s)\, F(s+y)dt + F(x+y) = 0,\quad 0\leq x\leq y
< \ue,
\end{equation}
where the function $F$ is expressed via the scattering data as follows:
$$F(x) = \frac{1}{2\pi} \int_{-\ue}^\ue [1-S(k)] e^{ikx}dk +\sum_{j=1}^J
s_je^{-k_jx},$$
and then the potential is found by the formula
 $$q(x) = - 2\, \frac{d A(x,x)}{d x}.$$ 
If one knows that $q =
0$ for $x\geq a$, then any of the data $\{ S(k)\}$, $\{ f(k)\}$,
$\da(k)$, $f'(k)$, $\fa k \geq 0$, where $\da(k)$ is the phase shift,
determine $q$ uniquely (\cite{R470}, p. 180). The phase shift $\da(k)$ is
defined by the relation $S(k)=e^{2i\da(k)}$.

In fact, a weaker
assumption was made in \cite{R470}: it was assumed that 
$\n q(x)\n \leq c_1 e^{-c_2 |x|^{\gamma}}$, where $\gm >
1$ and $c_1, c_2$ are positive constants. This assumptions implies that 
$f(k)$ 
is an entire function, $S(k)$ is meromorphic, and the only poles of
$S(k)$ in $\C_+ = \{ k:\, {\rm Im}\, k > 0\}$ are $ik_j$, $1 \leq j
\leq J$. Thus, $S(k)$ determines $J$ and $k_j$ uniquely, and $s_j = i\,
{\rm Res}_{k=ik_j}\, S(k)$ are also uniquely determined. Therefore, $q$ is
uniquely determined by $\{ S(k)\}_{k\geq 0}$. We refer the reader to
\cite{R470} for the uniqueness proof in the cases of other data.
\subsection{Potentials vanishing on half-line}\label{SS3.3} If the inverse scattering on the full line is
considered, then the scattering data are $$\Bigl\{  r(k)\Bigm\n_{\fa k
\geq 0}, \, k_j,\, s_j,\, 1 \leq j \leq J
\Bigr\},$$ 
where $r(k)$ is the reflection coefficient, $-k_j^2$ are the bound
states, and $s_j > 0$ are norming constants (see \cite{M} and \cite{R278}, 
p. 284). These data determine $q\in L_{1,1}(\R)$ uniquely. However, if
one knows that $q(x) = 0$ for $x < 0$, then the data $\{ r(k)\}_{\fa k
\geq 0}$ alone determine $q$ on $\R_+ = [0,\ue)$ uniquely \cite{R470}, p.
181. Indeed, if $q = 0$ on $\R_-$, then the scattering solution
on $\R$ is $u = e^{ikx} + r(k) e^{-kx}$ for $x < 0$ and $u=t(k)\,
f(x,k)$ for $x > 0$, where $t(k)$ is the transmission coefficient, which
is unknown, and $f(x,k)$ is the Jost solution. Thus
\begin{equation}
\label{E3.2}
\frac{ik[1-r(k)]}{1 + r(k)} = \frac{u'(-0,k)}{u(0,k)} =
\frac{u'(0,k)}{u(0,k)} = \frac{f'(0,k)}{f(0,k)}:= I(k).
\end{equation}
Therefore,  $r(k)$ determines uniquely the $I$-function $I(k)$. This 
function
determines $q$ uniquely \cite{R470}, p. 108. The proof in \cite{R470}
is based on $C_+$ property of the pair $\{ \ell_1,\ell_2\}$, $\ell_j:= -
\frac{d^2}{dx^2} + q_j(x)-k^2$. This property holds for $q_j\in L_{1,1}$
and says that the set $\{ f_1(x,k)\, f_2(x,k\}_{\fa k \geq 0}$ is
complete (total) in $L^1(\R_+)$ \cite{R470}, p. 104.
\subsection{Potentials known on a part of the interval}\label{SS3.4} Consider the equation $\ell u:= -u'' + q(x) u
= k^2 u$, $x\in [0,1]$, $u(0) = u(1) = 0$, $q = \ov{q}\in L^1([0,1])$.
Fix $0 < b \leq 1$. Assume $q$ on $[b,1]$ known and the subset $\{
\la_{m(n)}\}_{\fa n = 1,2,3\ldots}$ of the eigenvalues $\la_n= k^2_n$ of
$\ell $ is known, where $\frac{m(n)}{n} = \frac{1}{\si}\, (1 +
\ve_n)$, $\si = {\rm const} > 0$, $|\ve_n| < 1$,
$\sum^\ue_{n=1}\n \ve_n\n < \ue$.
\begin{theorem}[\cite{R387}, \cite{R460}, {\cite{R470}, p. 176}] If $\si
\geq 2b$, then the above data determine $q$ on $[0,b]$ uniquely.
\end{theorem}
For example, if $\si = 1$ and $b = \ha$, then the theorem says that $q$
is uniquely determined by one spectrum.
\section{Inverse scattering for inhomogeneous Schr\"odinger
equation}\label{S4}
Let $\ell u - k^2 u:= - u'' + q(x)u - k^2 u = \da(x)$, $x\in \R^1$,
$\lim_{\n x\n\ra \ue}(\frac{\pa u}{\pa \n x\n} - iku) = 0$. Assume $q =
\ov{q}$, $q = 0$ for $\n x\n > 1$, $q \in L^\ue[-1,1]$. Suppose the data $\{
u(-1,k),\, u(1,k)\}_{\fa k > 0}$ are given. It is proved in
\cite{R391} that these data determine $q(x) $ uniquely (see also 
\cite{R470}, p. 204).
\section{Krein's inversion method}\label{S5}
In \cite{R470}, p. 186, apparently for the first time, Krein's inversion
method (see \cite{K1}, \cite{K2}) was presented with detailed proofs, and 
it
was proved additionally that this method yields the unique potential which
reproduces the original scattering data $\ES$. For simplicity, let us
describe Krein's method assuming that there are no bound states, so that
the scattering data are $\{ S(k)\}_{\fa k \geq 0}$. Then Krein's method
can be described as follows: 

Given $S(k)$ with ${\rm ind}_\R S(k) = 0$,
where ${\rm ind}_\R S(k)$ is the index of $S(k)$, one finds $f(k)$ by the
formula 
$$f(k) = {\rm exp}\Big\{  \frac{1}{2\pi i} \int^\ue_{-\ue} \frac{\ln
S(-y)}{y-k}\, dy\Big\},\quad {\rm Im}\, k > 0,$$
then one calculates 
$$H(t) = \frac{1}{2\pi} \int^\ue_{-\ue}e^{-ikt}\big( \frac{1}{\n
f(k)\n^2}-1 \big) dk,$$
then one finds $\Gm_x(t,s)$ by solving the equations:
\begin{equation}
\label{E5.1}
\Gm_x(t,s) + \int_0^x H(t-u)\, \Gm_x(u,s)du = H(t-s),\quad 0 \leq t\,,
s \leq x.
\end{equation}
This equation is uniquely solvable if
\begin{equation}
\label{E5.2}
S(k) = \ov{S(-k)} = S^{-1}(k),\quad k\in \R;\quad {\rm ind}_\R\, S(k) = 0,
\end{equation}
and
\begin{equation}
\label{E5.3}
\ab F(x)\ab_{L^\ue(\R_+)} + \ab F(x)\ab_{L^1(\R_+)} + \ab x F'(x)\ab_{L^1(\R_+)}<\ue,
\end{equation}
where $F(x) = \frac{1}{2\pi} \int_{-\ue}^\ue [1-S(k)] e^{ikx}dk$.
>From $\Gm_x(t,s)$ one calculates $2\Gm_{2x}(2x,0):= a(x)$ and, finally,
$q(x) = a^2(x) + a'(x)$. In \cite{R470}, p. 197 it is proved that the
steps of the above inversion procedure are invertible, and the
constructed potential generates the original data $S(k)$ if conditions
(6.2)-(6.3) hold, the case when bound states are present is 
considered, and
the advantages of Krein's method for numerical implementation are
discussed.
\section{Invertibility of the steps in Gelfand-Levitan, Marchenko, and
Krein's methods}
The Gel'fand-Levitan (GL) method for finding $q(x)$ given the
corresponding spectral function $\rho(\la)$, consists of the following
steps:
\begin{equation}
\label{E6.1}
\rho \Ra L \Ra K \Ra q,
\end{equation}
where 
\begin{align*}
L(x,y) & = \int^\ue_{-\ue} \rho_0(x,\la)\, \rho_0(y,\la)d\si(\la),\quad
d\si = d(\rho-\rho_0),\\
d\rho_0 = \left\{ \begin{matrix} \frac{\sqrt{\la}\, d\la}{\pi}, & \la
\geq 0,\\
0, & \la < 0 ,
\end{matrix}\right.
\end{align*}
$K(x,y)$ is found from the GL equation
\begin{equation}\label{E6.2}
K(x,y) + \int_0^x K(x,s)\, L(s,y) ds + L(x,y) = 0,\quad 0\leq y \leq x,
\end{equation}
and the potential is found by the formula $q(x) = 2\frac{dK(x,x)}{dx}$.

Let us assume that the spectral functions $\rho(\la)$ satisfy two
assumptions:\\ $A_1)$  If $h\in L_0^2(\R_+)$ is arbitrary, $H(\la):=
\int_0^\ue h(x)\vp_0(x,\la)dx$, and
$\int^\ue_{-\ue}H^2(\la)d\rho(\la) = 0$, then $h = 0$. Here
$\vp_0(x,\la) = \frac{\sin(x\sqrt{\la)}}{\sqrt{\la}}$.\\
$A_2)$ If  $\int^\ue_{-\ue}\n H(\la)\n^2 d(\rho_1-\rho_2) = 0$ $\fa h\in
L_0^2(\R_+)$, then $\rho_1=\rho_2$.\vpar
These assumptions are satisfied, for example, for the operators 
$l_j=-\frac{d^2}{dx^2} + 
q_j(x)$ which are in the limit-point at infinity.
Under these assumptions it is proved in \cite{R470}, p. 128,  \cite{R460}, 
that each step in 
\eq{6.1} is invertible: 
$$\rho \Leftrightarrow L\Leftrightarrow K
\Leftrightarrow q$$ 
Methods for calculating $d\rho$ from $\ES$, the 
scattering data, and $\ES$ from $d\rho$, are given in \cite{R470}, p. 131,  
and the set of spectral functions, corresponding to $q\in H_{\rm loc}^m(\R_+)$
 is characterized: these are the $\rho$'s satisfying assumptions $A_1)$ 
and 
$A_3)$, where $A_1)$ is stated above and $A_3)$ is the following 
assumption: the function 
$L(x)\in H_{\rm loc}^{m+1}(\R_+)$, where 
$$L(x):=\int_{-\ue}^\ue
\frac{1-\cos(x\sqrt{\la})}{2\la}d(\rho-\rho_0),$$ so that 
$L(x)=L(\frac{x}{2},\frac{x}{2})$, where $L(x,y)$ is defined below \eq{6.1}.

The Marchenko $(M)$ inversion consists of the following steps:
$$\ES \Rightarrow F\Rightarrow A\Rightarrow q,$$
where $\ES:=\{ S(k), k_j, s_j, 1\le j\le J\}$ are the scattering data,  
$$F(x):=\frac{1}{2\pi}\int_{-\ue}^\ue[1-S(k)]e^{ikx}dk+\sum_{j=1}^J s_j 
e^{-k_jx},$$  $A=A(x,y)$ is the (unique) solution of the equation 
\eq{3.1},\, 
 Section \ref{SS3.2}, and 
$$q(x)=-2\frac{dA(x,x)}{dx}.$$ 
It is proved in 
\cite{R470}, p. 143,  that each step in (7.3) is invertible if $q\in 
L_{1,1}$. 
The scattering data, corresponding to $q\in L_{1,1}$, are characterized by the following conditions:
\begin{enumerate}
	\item[a)] 
	\[{\rm ind}_\R S(k)= \begin{cases}
	-2J \quad \text{if } f(0)\ne 0\\
	-2J-1\quad \text{if } f(0)=0,
	\end{cases}
\]
\item[b)] $k_j>0,s_j>0,1\le j\le J; S(k)=\overline{S(-k)} = S^{-1}(k), 
k\ge 0,S(\ue)=1$,
\item[c)] Equations (6.2)-(6.3) of Section 6 hold.
\end{enumerate}
Let $A(y):=A(0,y)$, where $A(x,y)$ is the transformation kernel 
defined in Section 4.2. It is proved in \cite{R470}, p. 147,  that $A(y)$ 
solves the equation:
\begin{equation}
	\label{E6.4}
	F(y)+A(y)+\int_{-\ue}^\ue A(t)F(t+y)dt=A(-y),\quad-\ue<y<\ue.
\end{equation}
If conditions a) -- c) hold and $k\f|f(k)|^2-1\g\in L^2(\R_+)$, then $q\in 
L^2(\R_+)$.

We have mentioned invertibility of the steps in Krein's inversion method
in Section 5. The proof of these results is given in \cite{R470}.
\section{The
Newton-Sabatier and Cox-Thompson procedures are not inversion
methods.} The Newton-Sabatier and Cox-Thompson procedures (\cite{N},
\cite{CS}, \cite{CT}) for finding $q$ from the set of fixed-energy phase
shifts are not inversion methods: it is not possible, in general, to carry
these procedures through, and  it is not proved that if these procedures 
can be
carried through, then the obtained potential  reproduces the 
original phase
shifts. These procedures are fundamentally wrong because their basic
assumptions are wrong: the integral equation, used in these procedures, in
general, is not uniquely solvable for some $r>0$, and then the procedures
break down. A detailed analysis of the Newton-Sabatier procedure is given
in \cite{R431}, \cite{R445}, \cite{R470}, p. 166,  and a counterexample to
the uniqueness claim in \cite{CT} is given in \cite{R433}. 

\section{Resonances}
If $q\in Q_a$, then the numbers $k\in \C_-=\{ z:\Im z\leq 0\}$, for which 
the 
equation $u+T(k)u=0$ has non-trivial solutions, are called resonances. 
Here
	\[
	T(k)u=\int_{B_a} g(x,y,k)q(y)u(y)dy, \quad
g(x,y,k):=\frac{e^{ik|x-y|}}{4\pi|x-y|}. \] In the one-dimensional case
resonances are zeros of the Jost function $f(k)$ in $\C_-$. It was proved
in \cite{R19} that if $q\in Q_a\cap C^1(B_a)$ then there are no resonances
in the region $\Im k>c-b \ln|k|$ for large $|k|$, where $b>0$ and $c$
are constants. In \cite {R74}, \cite{R125}, \cite{R128}, \cite{R130},
\cite{R145}, \cite{R146}, \cite{R148}, \cite{R149}, \cite{R175},
\cite{R178}, \cite{R190} methods for calculating resonances, perturbation
theory for resonances, variational principles for resonances, asymptotics
of resonant states, and the relation to eigenmode and singularity
expansion methods are given. In \cite{R165} existence of infinitely many
purely imaginary resonances is proved for $q\in Q_a$. In \cite{R278}, pp.
278--283, the following results are proved: \begin{enumerate}
	\item If $q\in L_{1,1}$ then $q=0$ for $x>2a$ if and only if 
the corresponding Jost function $f(k)$ is entire, of exponential type 
$\ge 2a$, bounded in $\C_+$ and $\lim_{|k|\to \ue,\,k\in \C_+} f(k)=1$;
	\item Let $Q_n:=\int_0^\ue x^n|q(x)|dx$. If $Q_n=O(n^{bn}), 0\le 
b<1$, and $q\not\equiv 0$, then $q$ generates infinitely many resonances;
	\item Let $a>0$ be arbitrarily small fixed number. There exists 
$q=\bar{q}\in C_0^\ue(0,a)$ which generates infinitely many purely imaginary 
resonances.
	
Note that any $q\in C_0^\ue(0,a),q=\bar{q}$, which does not change sign in an arbitrary small left neighborhood of $a$, i.e., on $(a-\da,a)$, where $\da>0$ is arbitrarily small, cannot produce infinitely many purely imaginary resonances.
\end{enumerate}
Example of application of the result (2): if $|q(x)|\le c_1 
e^{-c_2x^\gm},\,\, x\ge 0$, $\gm>1$, then $q$ generates infinitely many 
resonances. 
Indeed, if $t^\gm=y$ and $c:=c_2^{\frac{1}{\gm}}$, then
\begin{eqnarray*}
Q_n & = & c_1\int_0^\ue x^ne^{-c_2x^\gm}dx \le c_1\frac{1}{c^{n+1}}\int_0^\ue (cx)^ne^{-(cx)^\gm}dxc = \frac{c_1}{c}\frac{1}{c^n}\int_0^\ue t^n e^{-t^\gm}dt\\
&=& \frac{c_2}{c_n}\int_0^\ue y^{\frac{n+1}{\gm}-1} e^{-y}dy = 
c_2\frac{\Gm(\frac{n+1}{\gm})}{c^n}\le c_3 n^{bn},\,\,\, 
b=\frac{1}{\gm}>1.	
\end{eqnarray*}
Thus, if $q\in Q$, then it generates infinitely many resonances.
\section{Born inversion is always an ill-posed problem}
Born inversion has been quite popular among engineers and physicists. 
The exact scattering amplitude is:
\begin{equation}
	\label{E9.1}
	-4\pi A(\be,\al,k)=\int_D e^{-ik\be \cdot x} u(x,\al,k)q(x)dx,
\end{equation}
where $u(x,\al,k)$ is the scattering solution. In the Born approximation one 
replaces $u(x,\al,k)$ in \eq{9.1} by the incident field $u_0=
e^{ik\al\cdot x}$ and gets:
\begin{equation}
	\label{E9.2}
	-4\pi A(\be,\al,k)\approx \int_D e^{-ik(\be-\al)x} q(x)dx,
\end{equation} where the error of this formula is small if $q$ is small in
some sense, specified in \cite{R249}. If $A(\be,\al,k)$ is known for all
$k>0,\be,\al\in S^2$, then, for any $\xi\in\R^3$, one can find
(non-uniquely) $k,\be,\al$ so that $\xi=k(\al-\be)$, and \eq{9.2} gives
the equation $\tilde{q}(\xi)= \int_D e^{-i\xi\cdot x}q(x)dx$. In
\cite{R249} it was pointed out that although the direct scattering problem
can be solved in the Born approximation if $q$ is small quite accurately,
and the error estimate for this solution can be derived easily, the
inverse scattering problem in the Born approximation is always an
ill-posed problem, no matter how small $q$ is. In \cite{R249} and in
\cite{R470}, p. 307, this statement is explained and a stable inversion
scheme in the Born approximation is given with an error estimate, provided
that $q$ is small. It is also explained that collecting very accurate
scattering data is not a good idea if the inversion is done in the Born
approximation: in this approximation even the exact data have to be
considered as noisy data.

Let us state and prove just one Theorem 
 that makes the above easier to understand.
\begin{theorem}
If $q=\bar{q}$ and $|q(x)|\le c(1+|x|)^{-b},b>3$, then the equation
\begin{equation}
	\label{E9.3}
	-4\pi A(\be,\al,k)=\int_D e^{ik(\be-\al)\cdot x} q(x)dx
\end{equation}
implies $q=0$.
\end{theorem}

\pn
\textbf{{Proof:}} The exact scattering amplitude satisfies the relation (optical theorem):
\begin{equation}
	\label{E9.4}
	4\pi {\rm Im }A(\be,\be,k)=k\int_{S^2} |A(\be,\al,k)|^2d\al.
\end{equation}
>From \eq{9.3} and the assumption $q=\bar{q}$ it follows that ${\rm Im }A(\be,\be,k)=0$, so \eq{9.4} implies 
\begin{equation}
	\label{E9.4*}
	A(\be,\al,k)=0.
\end{equation}
If \eq{9.4*} holds for all $\be,\al\in S^2$ and $k>0$, then $q=0$. If 
$q\in Q_a$ and \eq{9.4*} holds for a {\it fixed} $k>0$ and all $\be,\al\in 
S^2$, 
then $q=0$ by the Ramm's uniqueness theorem from Section \ref{S2}.
\qed

\section{Inverse obstacle scattering with fixed-fre\-quen\-cy data.}\label{S10}
Let $D\subset \R^3$ be a bounded domain with a Lipschitz boundary $S$
(or less smooth boundary, a boundary with finite perimeter, see
\cite{R470}, pp. 227-234). Let $A(\be,\al)$ be the corresponding
scattering amplitude at a fixed $k > 0$. The boundary condition
on $S$, which we denote $\Gamma$,
is homogeneous, one of the three types: the Dirichlet $(\D)$: $u\rain{S}
= 0$, the Neumann $(\EN)\rain{S} = 0$, the Robin $(\ER)$: $(u_N +
h(s)u)\rain{S} = 0$. Here $N$ is the outer unit normal to $S$, $h(s)$ is
a piecewise-continuous bounded function, ${\rm Im}\, h \geq 0$. The last
inequality guarantees the uniqueness of the solution to the scattering
problem. This solution for nonsmooth $S$ is understood in the weak
sense. The inverse scattering problem consists of finding $S$ and the
boundary condition $(\D)$, $(\EN)$ or $(\ER)$ type, and the function $h$
if condition $(\ER)$ holds. The uniqueness of the solution to this
problem was first proved by the author (\cite{R190}) for Lyapunov
boundaries and then in \cite{R343} for boundaries with finite
perimeter. For such boundaries the normal is understood in the sense of
Federer \cite{F}, \cite{R470}, p. 227. A domain $D\subset \R^3$ has finite
perimeter if $S:= \pa D$ has finite Hausdorff $H_2(S)$ measure, i.e.,
$H_2(S) = \lim_{\ve \downarrow 0} \, \inf_{B_j}\sum_j r_j^2 < \ue$,
where the infimum is taken over all coverings of $S$ by open
two-dimensional balls of radii $r_i < \ve$. For domains with finite
perimeter Green's formula holds:
$$ \int_D \na u\, dx = \int_{\pa^* D}N(s)\, u(s)H_2(ds),$$
where $\partial^* D$ is the reduced boundary of $D$, i.e. the subset of 
points
of $\pa D$ at which the normal in the sense of Federer exists.
\vpar
It is proved in \cite{R343} that if $D_1$ and $D_2$ are two bounded
domains with finite perimeter, $A_j$ are the corresponding scattering
amplitudes, and $u_j(x,\al)$ are the corresponding scattering solutions,
then
\begin{equation}\label{E10.1}
 4\pi [A_1(\be,\al) - A_2(\be,\al)] = \int_{S_{12}} [u_1(s,-\be)
u_{2N}(s,\al)-u_{1N} (s,-\be)u_2(s,\al)] ds,\end{equation}
where  $S_{12}:= \pa D_{12}$, $D_{12}:= D_1 \cup D_2$, and we assume
$S_j$ Lipschitz for simplicity of the formulation of the result. If
$A_1=A_2$ $\fa \al,\be\in S^2$, then \eq{10.1} yields
\begin{equation}
\label{E10.2}
0 = \int_{S_{12}}\bka
u_1(s,-\be)u_{2N}(s,\al)-u_{1N}(s,-\be)u_2(s,\al)\bkz ds\quad \fa
\al,\be\in S^2.
\end{equation}
>From \eq{10.2} by the ''lifting`` method  \cite{R470}, p. 236, one derives
that
\begin{equation}
\label{E10.3}
G_1(x,y) = G_2(x,y)\quad \fa x,y\in D'_{12}:= \R^3\bs D_{12},
\end{equation}
where $G_j$, $j = 1,2$ are Green's functions for the operator $\na^2 +
k^2$ in domain $D_j$, corresponding to the boundary condition $\Gm_j$
$(\Gm_j= (\D), (\EN)$ or $(\ER)$). If $S_1 \neq S_2$ then \eq{10.3} leads
to a contradiction, which proves that $S_1=S_2$. Indeed, one takes a
point $x\in S_1\cap D'_2$, and let $y\ra x$. Since $x\not\in D_2$ one
has 
$$\bigl\n \Gm_1 G_2(x,y)\bigr\n \:\mathr{\ra}{y\ra x}\infty\quad 
\mbox{and}\quad
\n \Gm_1 G_1(x,y)\n = 0\quad \fa y\in D'_1.$$
This contradiction proves that $S_1=S_2:= S$, so $D_1=D'_2 = D$, $\Gm_1
= \Gm_2 := \Gm$.\\ 
If $\Gm_1(x,y) = 0$, $x\in S$, $y\in D'$, then $\Gm =
(\D)$.\\ 
If $G_N(x,y) = 0$, $x\in S$, $y\in D'$, then $\Gm = (\EN)$.\\ 
If
$\frac{G_N}{G}:= -h$ on $S$, then $\Gm = (\ER)$ and $h$ on $S$ is
recovered.
\vpar
In \cite{R470}, p. 237, a more complicated problem is treated: let
\begin{equation}
\label{E10.4}
Lu:= \na\cdot \f a(x)\na u\g + q(x) u = 0\quad \mbox{in }\R^3,
\end{equation}
where 
\begin{align}
\label{E10.5}
& \mbox{$a(x) = a^+ $ in $D$, $a(x) = {a}^- := a_0$ in $D':= \R^3\bs D$,
$q(x) = k^2 a^+ := q^+ $ in $D$,}
\\
\label{E10.6}
& q(x)= k_0^2 a_0:= q^- \mbox{ in } D',\quad a^\pm,q^\pm,k_0,k > 0 \mbox{
are constants.}
\\
\label{E10.7}
& a^+ u^+_N = a_0 u_N^- \mbox{ on } S,\quad a^+ \neq a^-,
\\
\label{E10.8}
& u = u_0 + A(\be,\al,k_0)\, \frac{e^{ik_0r}}{r} + o\f \frac{1}{r}\g ,
\quad r:= \n x\n \ra \ue,\, u_0:= e^{ik_0\al\cdot x}.
\end{align}
Assume $k_0 > 0$ is fixed, denote $A(\be,\al,k_0):= A(\be,\al)$.
\begin{theorem}[{\cite{R470}, p. 237}]
The data $A(\be,\al)$ $\fa
\be,\al\in S^2$, $a_0$ and $k_0$, determine $S,a^+,$ and $k$ uniquely.
\end{theorem}
The proof is based on the result in \cite{R371} on the behavior of
fundamental solutions to elliptic equations with discontinuous senior
coefficients. In \cite{R325} stability estimates are obtained for the
recovery of $S$ from the scattering data and an inversion formula is
given. Let us formulate these results. Consider two star-shaped
obstacles $D_j$ with boundaries $S_j$ which are described by the
equations $r = f_j(\al)$, $r = \n x\n$, $\al = \frac{x}{\n x\n}\,$, $j =
1,2$. Assume that $0 < c \leq f_j(\al) \leq C$, $\fa \al\in S^2$, and
$S_j\in C^{2,\la}$, $0<\la \leq 1$, with $C^{2,\la}$-norm of the
functions, representing the boundaries $S_j$ bounded by a fixed constant. 
Let 
$$\rho:= \max \{ \sup_{x\in S_1}
\inf_{y\in S_2}\, \n x-y\n, \: \sup_{y\in S_1}\, \inf_{x\in S_1}\, \n
x-y\n\}$$
be the Hausdorff distance between the obstacles, and
$$\sup_{\be,\al}\n A_1(\be,\al)-A_2(\be,\al)\n < \da,\quad k > 0$$
is fixed, and the Dirichlet boundary condition holds on $S_1$ and $S_2$.

In \cite{R325}, \cite{R470}, p. 240, the following results are proved:
\begin{theorem} 
Under the above assumptions one has 
$$\rho \leq c_1 \f \frac{\log\n \log
\da\n}{\n \log \da\n}\g^{c_2},$$ 
where $c_1$ and $c_2$ are positive
constants independent of $\da$.
\end{theorem}
\begin{theorem} 
There exists a function $\nu_\eta(\al,\ta)$ such that
\begin{equation}
\label{E10.9}
-4\pi \lim_{\eta\ra 0}\int_{S_2} A(\ta',\al)\, \nu_\eta(\al,\ta)d\al = -
\frac{\n \xi\n^2\, \s{\chi}_D(\xi)}{2}\,,
\end{equation}
where 
$$\mbox{$\s{\chi}_D(\xi):= \int_D e^{-i\xi\cdot x}\, dx$, \quad 
$\ta',\ta\in
M_k$, $k(\ta'-\ta) = \xi\in \R^3$,}$$ 
$M_k:= \{ \ta:\, \ta\in \C^3,\: \sum^3_{j=1}\ta_j^2 = k^2\}$, $k > 0$ is
fixed, and the equation for the scattering solution is
\begin{align}
\label{E10.10}
&(\na^2+k^2) u  = 0\quad \mbox{in } D':= \R^3 \setminus D,u\n_S = 0,\\
& u = e^{ik\al\cdot x} + A(\be,\al)\, \frac{e^{ikr}}{r} +
o(\frac{1}{r}),\quad r := \n x\n\ra \ue,\: \be= \frac{x}{r}\,,\: \al\in
S^2. \label{E10.11}
\end{align}
\end{theorem}

It is an {\bf open problem} to find an algorithm for calculating
$\nu_\eta(\al,\ta)$ given $A(\be,\al)$.

In the literature there is a large number of papers containing various
methods for finding $S$ given $A(\be,\al)$. In \cite{R470}, pp. 245-253 there
are some comments on these methods, The basic points of these comments
are:  most of these methods are not satisfactory, there are no error 
estimates
for these methods, no guaranteed accuracy for the solution of the
inverse obstacle scattering problem are currently available. The
published numerical results with good agreement between the original
obstacle and its reconstruction are obtained because the original
obstacle was known a priori. Thus, it is still an open problem to
develop a stable inversion method for finding $S$ from $A(\be,\al)$ and
to give an error estimate for such a method. In \cite{R190}, p. 94 (see
also \cite{R172}, \cite{R173}, \cite{R278}, pp. 126-130, for strictly
convex obstacles an analytic formula for calculating $S$ from
high-frequency data $A(\be,\al,k)\rain{k\ra \ue}$ is given, and the
error estimate for the $S$ recovered from this formula when the noisy
data $A_\da(\be,\be,\al,k)$ are used is also given.
\section{Inverse scattering with data at a  fixed-energy and a fixed
incident direction}
Let us pose the following inverse scattering problem:\\
ISP: {\em Given an arbitrary function $f(\be)\in L^2(S^2)$, an arbitrary
small number $\ve > 0$, a fixed $k > 0$, and a fixed incident direction
$\al\in S^2$, can one find a potential $q\in L^2(D)$ such that the
corresponding scattering amplitude $A_q(\be,\al,k):= A(\be)$ satisfies
the inequality
\begin{equation}
\label{E11.1}
\ab f(\be)-A(\be\ab < \ve,
\end{equation}
where $\ab {\cdot}\ab = \ab {\cdot}\ab_{L^2(S^2)}$, $D\subset \R^3$ is a
bounded domain, $D\subset B_a:= \{ x:\, \n x\n \leq a\}$?}
\vpar
Let us prove that the answer is yes, that there are infinitely many such
$q$ and give a method for finding such a $q$. The idea of the solution
of this ISP can be outlined as follows: start with the known formula
\begin{equation}
\label{E11.2}
-4\pi A(\be) = \int_D e^{-ik\be\cdot x}\, u(x)q(x)dx,
\end{equation}
where $u(x):= u(x,\al,k):= u_q$ is the scattering solution:
\begin{equation}
\label{E11.3}
[\na^2 + k^2 -q(x)]u = 0\quad \mbox{in }\R^3,
\end{equation}
$u$ satisfies \eq{10.8}, $\al\in S^2$ and $k > 0$ are fixed. Denote
\begin{equation}
\label{E11.4}
h(x):= h_q:= u(x)q(x),
\end{equation}
so
\begin{equation}
\label{E11.5}
A(\be) = -\frac{1}{4\pi}\int_D e^{-ik\be\cdot x}h_q(x)dx.
\end{equation}
{\bf Step 1.} Given $f(\be)$ and $\ve > 0$, find an $h\in L^2(D)$ such
that
\begin{equation}
\label{E11.6}
\ab f(\be) + \frac{1}{4\pi}\int_D e^{-ik\be\cdot x}h(x)dx\ab < \ve.
\end{equation}
There are infinitely many such $h$. Existence of such $h$ follows from 
Lemma 1.
\begin{lemma}
The set $\{ \int_D e^{-ik\be\cdot x} h(x)dx\}_{\fa h\in L^2(D)}$ is
total (complete) in $L^2(S^2)$.
\end{lemma}
In \cite{R517} some analytical formulas are given for calculating an $h$
satisfying \eq{11.6}.
\vpn {\bf Step 2.} Given $h\in L^2(D)$, find a $q\in L^2(D)$ such that
\begin{equation}
\label{E11.7}
\ab h-q(x)u_q(x)\ab_{L^2(D)}<\ve.
\end{equation}
This is possible because of  Lemma 2.
\begin{lemma}
The set $\{ q(x)u_q(x)\}_{\fa q\in L^2(D)}$ is total in $L^2(D)$. 
\end{lemma}
Let an
arbitrary $h\in L^2(D)$ be given. Consider the function
\begin{equation}
\label{E11.8}
q(x):= \frac{h(x)}{u_0(x)-\int_D g(x,y)h(y)dy}:=
\frac{h(x)}{\psi(x)},\quad g:= \frac{e^{ik\n x-y\n}}{4\pi \n x-y\n}\,.
\end{equation}
The scattering solution $u=q_q$ solves the equation
\begin{equation}
\label{E11.9}
u(x) = u_0(x)-\int_D g(x,y)\, q(y)\, u(y)\, dy,\quad u_0:= e^{ik\al\cdot 
x}.
\end{equation}
If the function $q(x)$, defined in \eq{11.8}, belongs to $L^2(D)$,
then define
\begin{equation}
\label{E11.10}
u_q(x):= u_0(x)-\int_D g(x,y) h(y)dy.
\end{equation}
The function 
\eq{11.10} satisfies equation \eq{11.9} because formula \eq{11.8}
implies $q(x) u_q(x) = h(x)$. Thus, the function \eq{11.10} is a
scattering solution. The corresponding scattering amplitude 
\begin{equation}
\label{E11.11}
A_q(\be) = - \frac{1}{4\pi} \int_D e^{-ik\be\cdot y}\, h(y)dy.
\end{equation}
Thus, inequality \eq{11.6} is satisfied, so formula \eq{11.8} gives a
solution to ISP.
\vpar
Suppose now that the function $q(x)$, defined by \eq{11.8}, does not
belong to $L^2(D)$. This can happen because the denominator $\psi$ in
\eq{11.8} may vanish. Note that small in $L^2(D)$-norm perturbations of
$h$ preserve inequality \eq{11.6}, possibly with $2\ve$ in place of
$\ve$. Since the set of polynomials is total in $L^2(D)$, one may assume
without loss of generality that $h(x)$ is a polynomial. In this case the
function $\psi(x)$ is infinitely smooth in $D$ (even analytic). Consider
the set of its zeros $N:= \{ x:\, \psi(x) = 0,\, x\in D\}$. Let $\psi:=
\psi_1 + i\psi_2$, where $\psi_1= {\rm Re}\, \psi$, $\psi_2 = {\rm Im}\,
\psi$. Small perturbations of $h$ allow one to have $\psi$ such that
vectors $\na \psi_1$ and $\na \psi_2$ are linearly independent on $N$,
which we assume. In this case $N$ is a smooth curve $\CE$, defined as
the intersection of two smooth surfaces:
\begin{equation}
\label{E11.12}
\CE:= \{ x:\, \psi_1(x) = 0,\quad \psi_2(x) = 0,\: x\in D\}.
\end{equation}
Let us prove that for an arbitrary small $\da > 0$ there is a $h_\da$,
$\ab h_\da - h\ab_{L^2(D)}<\da$, such that $q_\da\in L^\ue(D)$, where
\begin{equation}
\label{E11.13}
q_\da(x):= \left\{
\begin{array}{ll}
\frac{h_\da(x)}{u_0(x)-\int_D g(x,y) h_\da(y)dy} & \mbox{ in }D_\da,\ek
0 & \mbox{ in }N_\da .
\end{array}
\right.
\end{equation}
Here $D_\da := D\bs N_\da$, $N_\da : = \{ x:\, \n \psi(x)\n \geq \da >
0\}$. The function \eq{11.13} solves our inverse scattering problem ISP 
because the inequality $\ab
h_\da-h\ab_{L^2(D)} < \da$ for sufficiently small $\da$ implies the
inequality
\begin{equation}
\label{E11.14}
\ab f(\be) + \frac{1}{4\pi}\int_D e^{-ik\be\cdot x} \, h_\da(x)dx\ab < 2\ve.
\end{equation}
Let us now prove the existence of $h_\da$, $\ab h_\da -
h\ab_{L^2(D)}<\da$, such that $q_\da\in L^\ue(D)$, where $q_\da$ is
defined in \eq{11.13}. The $h_\da$ is given by the formula:
\begin{equation}
\label{E11.15}
h_\da(x)= \begin{cases}
0 & \mbox{ in }N_\da,\\
h & \mbox{ in }D_\da.
\end{cases}
\end{equation}
To prove this, it is sufficient to prove that 
$$\psi_\da:= u(x)-\int_D g(x,y) h_\da(y)dy$$ 
satisfies the inequality
\begin{equation}
\label{E11.16}
\min_{x\in D_\da}\n \psi_\da(x)\n \geq c(\da) > 0,
\end{equation}
because $h_\da(x)$ is a bounded function in $D$. 

To prove \eq{11.16} use
the triangle inequality:
\begin{eqnarray}
\nn \n \psi_\da(x)\n  &\geq & \Big\n u_0(x)-\int_D g(x,y) h(y)dy\Big\n 
- \int_D \n g(x,y)\n\: \n h(y)-h_\da(y)\n dy\\
\label{E11.17}
& \geq & \da - c\int_{N_\da} \frac{dy}{\n x-y\n}\,,\quad x\in D_\da,
\end{eqnarray}
where $c>0$ is a constant independent of $\da$, 
$$c=\frac{1}{4\pi}\,\max_{y\in N_\da}\n h(y)\n \leq \frac{1}{4\pi}\,
\max_{y\in D}\n h(y)\n.$$ 
Let us prove that
\begin{equation}
\label{E11.18}
I_\da:= \int_{N_\da} \frac{dy}{\n x-y\n} \leq c\da^2\n \ln \da\n\,,\quad
x\in D_\da,\quad \da\ra 0,
\end{equation}
where $ s > 0$ stands for various constants independent of $\da$. If
\eq{11.18} is proved, then \eq{11.16} is established, and this proves
that $q_\da(x)$ solves the ISP. To prove \eq{11.18} choose the new
coordinates $s_1,s_2,s_3,$ such that $s_1 = \psi_1(x)$, $s_2 = \psi_2(x)$,
$s_3 = x_3$, where the origin $O$ is on $N$, i.e., on the curve $\CE$,
$x_1$ and $x_2$ axes are in the plane orthogonal to
the curve $\CE$ at the origin.
This plane contains vectors $\na \psi_1$ and $\na \psi_2$ calculated at
the origin, and $x_3$ axis is directed along the vector product $[\na 
\psi_1,\na
\psi_2]$. The set $N_\da$ is a tubular neighborhood of $\CE$, and we
consider a part of this neighborhood near the origin. The set $N_\da$ is
a union of similar parts and for each of them the argument is the same.
Consider the Jacobian $\mathcal{J}$ of the transformation
$(x_1,x_2,x_3)\ra (s_1,s_2,s_3)$:
\begin{equation}
\label{E11.19}
\J= \frac{\pa(\psi_1,\psi_2,\psi_3)}{\pa(x_1,x_2,x_3)}= \begin{vmatrix}
\psi_{1,1} & \psi_{1,2} & \psi_{1,3}\\
\psi_{2,1} & \psi_{2,2} & \psi_{2,3}\\
0 & 0 & 1
\end{vmatrix} \neq 0,
\end{equation}
where $\psi_{i,j}:= \frac{\pa \psi _i}{\pa x_j}$, and we have used the
assumption that the vectors $\na \psi_1$ and $\na \psi_2$ are linearly
independent, so that 
$$ \begin{vmatrix} \psi_{1,1}  & \psi_{1,2}\\
\psi_{2,1} & \psi_{2,2}
\end{vmatrix} \neq 0$$ 
in our coordinates. Thus
\begin{equation}
\label{E11.20}
\n f\n + \n f^{-1}\n\leq c\quad \mbox{on }N,
\end{equation}
and, by continuity, in $N_\da$ for a small $\da$. The integral
\eq{11.18} can be written in the new coordinates as
\begin{equation}
\label{E11.21}
I_\da \leq \int_{\substack{\psi_1^2 + \psi_2^2 \leq \da^2\\ 0\leq s_3
\leq c_3}} \frac{dy}{\n y\n}\leq c \int_0^\da d\rho\, \rho \int_0^{c_3}
\frac{ds_3}{\sqrt{\rho^2 + s_3^2}}\,,
\end{equation}
where we have used   the estimate \eq{11.12} and \eq{11.20} and the 
following estimate:
\begin{equation}
\label{E11.22}
c_1(\psi_1^2 + \psi_2^2 + y_3^2)\leq \n y\n^2\leq c_2(\psi_1^2 +
\psi_2^2 + y_3^2)\quad \mbox{in }N_\da.
\end{equation}
We have:
$$\int_0^{c_3} \frac{ds_3}{\sqrt{\rho^2 + s_3^2}} = \ln(s_3 + \sqrt{p^2
+ s_3^2})\Big\n^{c_3}_0 \leq c\, \ln \frac{1}{\rho}\leq c\, \ln
\frac{1}{\da}\,.$$
Thus, 
$$I_\da \leq c\da^2 \n \ln \da\n,$$ and \eq{11.16} is verified.
\vpar
Let $\phi\in H_0^2(D)$, where $H_0^2(D)$ is the closure of $C_0^\ue(D)$
functions in the norm of the Sobolev space $H^2(D)$. If one replaces $h$
by $h + (\na^2 + k^2)\phi$ in \eq{11.11}, then $A_q(\be)$ remains
unchanged because 
$$\int_D e^{-ik\beta \cdot x}(\nabla^2+k^2)\phi(x) dx=0$$
for $\phi\in H_0^2(D)$.
 
One has
\begin{equation}
\label{E11.23}
\int_D g(x,y) [h(y) + (\na^2 + k^2)\phi]dy = \int_D g(x,y) h(y)dy = \phi(x).
\end{equation}
Thus, the potential
\begin{equation}
\label{E11.24}
q_\phi(x) := \frac{h(x) + (\na^2 + k^2)\phi(x)}{u_0(x)-\int_D g(x,y) h(y)
dy + \phi(x)}
\end{equation}
generates the same scattering amplitude as the potential \eq{11.8} for
any $\phi\in H_0^2(D)$. Choosing a suitable $\phi$, one can get a
potential with a desired property. Let us prove that $\phi\in H_0^2(D)$
can be always chosen so that the potential $q_\phi(x)$ has the property
${\rm Im}\, q_\phi \leq 0$, which physically corresponds to an absorption
of the energy.
\par Denote $L:=\na^2 + k^2$, $\phi_1 = {\rm Re}\, \phi$, $\phi_2:= 
{\rm
Im}\, \phi$, 
$$\psi:= u_0-\int_D g(x,y) h(y)dy = \psi_1 + i \psi_2,$$
$\psi_1:= {\rm Re}\, \psi$, $\psi_2:= {\rm Im}\, \psi$. Then
\begin{equation}
\label{E11.25}
q_\phi = \frac{[h_1 + L\phi_1 + i(h_2 + L\phi_2][\psi_1 + \phi_1 -
i(\psi_2 + \phi_2)]}{\n \psi + \phi\n^2}\,.
\end{equation}
Thus, ${\rm Im}\, q_\phi\leq 0$ if and only if 
\begin{equation}
\label{E11.26}
-(h_1 + L\phi_1)(\psi_2 + \phi_2) + (h_2 + L\phi_2)(\psi_1 + \phi_1)
\leq 0.
\end{equation}
Choose $\phi_1$ and $\phi_2$ so that
\begin{equation}
\label{E11.27}
L\phi_1 + h_1 = \psi_2 + \phi_2,\quad L\phi_2 + h_2 = -(\psi_1 + \phi_1)
\end{equation}
in $D$. Eliminate $\phi_2$ and get
\begin{equation}
\label{E11.28}
L^2\phi_1 + \phi_1 + Lh_1 - L\psi_2 + h_2 + \psi_1 = 0\quad \mbox{in } D.
\end{equation}
The operator $L^2 + I$ is elliptic, positive definite, of order four,
with boundary conditions
\begin{equation}
\label{E11.29}
\phi_1 = \phi_{1N} = 0\quad \mbox{on } S,
\end{equation}
because $\phi_1\in H_0^2(D)$. Therefore problems \eq{11.28} - \eq{11.29}
have a unique  solution $\phi_1\in H_0^2(D)$.
If $\phi_1$ solves \eq{11.28} -- \eq{11.29}, then
\begin{equation}
\label{E11.30}
\phi_2 := L\phi_1 + h_1-\psi_2
\end{equation}
solves the second equation \eq{11.27}. Function \eq{11.30} belongs to
$H_0^2(D)$. Indeed, it is the unique solution in $H_0^2(D)$ of the
equation
\begin{equation}
\label{E11.31}
L^2\phi_2 + \phi_2 + Lh_2 + L\psi_1 - h_1 + \psi_2 = 0\quad \mbox{in } D.
\end{equation}
One may ask if $\phi$ can be chosen so that ${\rm Im}\, q_\phi =
0$. A sufficient condition for this is the following one:
\begin{equation}
\label{E11.32}
(h_1 + L\phi_1)(\psi_2 + \phi_2) = (h_2 + L\phi_2)(\psi_1 + \phi_1).
\end{equation}
There are $\phi_1$ and $\phi_2$ in $H_0^1(D)$ such that \eq{11.32} is
satisfied. For example, $\phi_1$ and $\phi_2$ can be found from the
equations
\begin{align}
L\phi_1 + h_1 & =  \psi_1 + \phi_1, & \hspace{-2cm} \phi_1 \in H_0^1(D),\label{E11.33} \\
L\phi_2 + h_2 & = \psi_2 + \phi_2, & \hspace{-2cm}\phi_2 \in H_0^1(D),\label{11.34}
\end{align}
provided that $k^2 + 1$ is not a Dirichlet eigenvalue of the Laplacian in
$D$. This can be assumed without loss of generality because if $k^2 + 1$
is such an eigenvalue, then it will not be such an eigenvalue in $D_\da$
for a small $\da > 0$ (see \cite{R190}). However, this argument leaves
open the existence of $\phi_1,\phi\in H_0^2(D)$ for which \eq{11.32}
holds.
\section{Creating materials with desired refraction coefficient}
Let $D\subset \R^3$ be a bounded domain filled with a material with
known refraction coefficient $n_0^2(x)$, so that the wave scattering
problem is described by the equations
\begin{equation}
\label{E12.1}
L_0 u_0:= \bka \na^2 + k^2 n_0^2(x)\bkz \, u_0(x,\al) = 0\quad \mbox{in
}\R^3,
\end{equation}
\begin{equation}
\label{E12.2}
u_0(x,\al) = e^{ik\al\cdot x} + A_0(\be,\al)\, \frac{e^{ikr}}{r} + o\f
\frac{1}{r}\g,\quad r:= \n x\n \ra \ue,\: \be := \frac{x}{r}.
\end{equation}
Here $k > 0$ is fixed, $n_0^2(x)$ is a bounded function, such that 
$$\max_{x\in \R^3}\n n_0(x)\n = n_0 < \ue, \quad {\rm Im}\, n_0(x) \geq 
0,$$ 
so that absorption is possible, and
\begin{equation}
\label{E12.3}
n_0^2(x) = 1\quad \mbox{in } D':= \R^3\bs D.
\end{equation}
The question is: 

{\it Is it possible to create a material in
$D$ with a desired refraction coefficient $n^2(x)$ by embedding small
particles into $D$?} 

If yes, what is the number $\EN(\Da)$ of the
particles of characteristic size $a$ that should be embedded in a small
cube $\Da\subset D$, centered at a point $x\in D$, and what should be
the properties of these small particles?
\par 
A positive answer to the above question is given in \cite{R529} and this
answer requires to solve a many-body wave scattering problem for small
particles embedded in a medium. The theory was presented in \cite{R518},
\cite{R524}, \cite{R525}, \cite{R476}.
\par
One of the results can be stated as follows. Assume that the small
bodies $D_m$, $1 \leq m \leq M$, are all balls of radius $a$ and that
the following limit exists:
\begin{equation}
\label{E12.4}
\lim_{a\ra 0} a\EN(\s{D}) = \int_{\s{D}}N(x)dx
\end{equation}
for any subdomain $\s{D}\subset D$, where $N(x)\geq 0$ is a continuous
function in $D$, $N(x) = 0$ in $D'$.
\par
Let us consider the scattering problem:
\begin{align}
\label{E12.5} & L_0 u_a = 0\quad \mbox{in } \R^3\bs \bigcup^M_{m=1} D_m,
\\ \label{E12.6} & u_{aN} = \zeta_m u_a \quad \mbox{on } S_m:= \pa D_m,
\\ \label{E12.7} & u_a = u_0 + A_a(\be,\al)\, \frac{e^{ikr}}{r} + o\f
\frac{1}{r}\g,\quad r:= \n x\n \ra \ue,\: \frac{x}{r} = \be,
\end{align}
and assume that
\begin{equation}
\label{E12.8}
\zeta_m = \frac{h(x_m)}{a}\,,
\end{equation}
where $x_m$ is the center of the ball $D_m$, and $h(x)$, ${\rm Im}\,
h(x) \leq 0$, is an a priori given arbitrary continuous in $D$ function,
$n(x) = 0$ in $D'$, $x_m \ra x$ as $a\ra 0$.
\par
Finally assume that
\begin{equation}
\label{E12.9}
d=O(a^{1/3}),\quad a \ra 0,
\end{equation}
where $d$ is the smallest distance between two distinct particles
(balls).
\begin{theorem}
Under the above assumptions there exists the following limit:
$$\lim_{a\ra 0}u_a(x) = u(x).$$ 
This limit solves the scattering problem 
\begin{align}
\label{E12.10} & \bka \na^2 + k^2 n^2(x)\bkz u = 0\quad \mbox{in }\R^3,
\\ \label{E12.11}
& u(x) = u_0(x) + A(\be,\al)\, \frac{e^{ikr}}{r} + o\f
\frac{1}{r}\g,\quad r=\n x\n \ra \ue,\quad \frac{x}{r} = \be,
\end{align}
where
\begin{equation}
\label{E12.12}
n^2(x) = n_0^2(x)-k^{-2} p(x) \mbox{ in }D,\quad n^2(x)=1 \mbox{ in }D',
\end{equation}
where
\begin{equation}
\label{E12.13}
p(x):= \frac{4\pi\, N(x)\, h(x)}{1 + h(x)}\,.
\end{equation}
\end{theorem}
\begin{corollary}
Given $n_0^2$ and an arbitrary continuous function $n^2(x)$ in $D$,
${\rm Im} \, n^2(x)\leq 0$, one can find (non-uniquely) three functions
$N(x)\geq 0$, $h_2:= {\rm Im}\, h(x)\leq 0$, and $h_1(x) = {\rm Re} \,
h(x)$, such that \eq{12.12} holds with $p(x)$ defined in \eq{12.13}.
\end{corollary}

The result of Theorem 6 was generalized to the case of small particles of 
arbitrary shapes in \cite{R529}.


\begin{thebibliography}{99}



\bibitem{CS} 
K. Chadan and T. Sabatier,  {\em Inverse problems in quantum
scattering theory}, Springer Verlag, New York, 1989.

\bibitem{CT} 
J. Cox and K. Thompson, Note on the uniqueness of the
solution of an equation of interest in inverse scattering
problem,  {\em J.~ Math.~Phys.}, {\bf 11 (3)}, (1970), 815-817.


\bibitem{F} H. Federer,  
{\em Geometric Measure Theory},
 Springer Verlag, New York, 1969.

\bibitem{H} 
M. Horvath,  Inverse scattering with fixed energy and an
 inverse eigenvalue problem on the  half-line, {\em Trans. Amer. Math. 
Soc.},
{\bf 358}, (11)(2006), 5161-5177.

\bibitem{K1} 
M.G. Krein,  Theory of accelerants and $S$-matrices of canonical
differential systems, {\em Doklady Acad.~Sci.~ USSR}, {\bf 111}, N6,
(1956), 1167-1170.

\bibitem{K2} 
 M.G. Krein,
{\em Topics in differential and integral equations and
operator theory}, {Birkh\"auser}, Basel, 1983.

\bibitem{L} 
B. Levitan, {\it Inverse Sturm-Liouville problems}, VNU Press,
Utrecht, 1987.


\bibitem{M} 
V. A. Marchenko, {\it  Sturm-Liouville operators and applications},
Birkh\"auser, Basel, 1986.


\bibitem{N} R. Newton, 
{\it Scattering Theory of Waves and Particles},
Springer, New York, 1982.


\bibitem{R3} 
A.G. Ramm,
 Investigation of the scattering problem in
some
 domains with infinite boundaries I, II, {\em Vestnik Leningradskogo
Universiteta, Ser. Math., Mechan. and Astronomy},  
{\bf 7}, (1963), 45-66;  19, (1963), 67-76. 

\bibitem{R19}     
A.G. Ramm,
Domain free from  the resonances in
the three-dimensional scattering problem, {\em Doklady  Acad of Sci. 
USSR}, {\bf 166}, 
(1966), 1319-1322. 

\bibitem{R74}     
A.G. Ramm,
Calculation of the quasistationary states
in
 quantum mechanics, {\em Doklady Acad. Sci. USSR}, {\bf 204}, (1972), 
1071-1074.

\bibitem{R125}   
A.G. Ramm,
Nonselfadjoint operators in diffraction
and
 scattering, {\em Math.Methods in appl.sci.}, {\bf 2}, (1980),
 327-346.
\bibitem{R128}  
A.G. Ramm,
Theoretical and practical aspects of
singularity
 and eigenmode expansion methods, 
 {\em IEEE A-P}, {\bf 28}, N6, (1980), 897-901.
\bibitem{R130}   
A.G. Ramm,
A variational principle for resonances, {\em J. Math.
 Phys.}, {\bf 21}, (1980), 2052-53.

\bibitem{R145}     
A.G. Ramm,
Mathematical foundations of the
singularity and
 eigenmode expansion methods, {\em J. Math. Anal. Appl.},
{\bf 86}, (1982), 562-591.
\bibitem{R146}  
A.G. Ramm, P.A.Mishnaevsky,
Asymptotics of resonant states, {\em J.Math.Anal.Appl.}, {\bf 87}, (1982),
 323-331.
\bibitem{R148}     
A.G. Ramm,
Variational principles for resonances II, {\em
J. Math. Phys.}, {\bf 23}, N6, (1982), 1112-1114.
\bibitem{R149}    
A.G. Ramm,
Perturbation of resonances. {\em J. Math. Anal.
Appl.}
{\bf 88}, (1982), 1-7.
 
\bibitem{R165}   
A.G. Ramm,
Existence of infinitely many purely
imaginary
 resonances in the problem of potential scattering.
 {\em Phys. Lett.}, {\bf 101A}, N4, (1984), 187-188.

\bibitem{R172}  
A.G. Ramm,
On inverse diffraction problem. {\em J. Math.
Anal.
 Appl.}, {\bf 103}, (1984), 139-147.
 
\bibitem{R173}  
A.G. Ramm,
Inverse diffraction problem, Inverse
methods in electromagnetic imaging,  Reidel, Dordrecht, 1985, pp. 231-249.  
( ed. W. Boerner)

\bibitem{R175} 
A.G. Ramm,
Extraction of resonances from transient
fields, {\em IEEE AP Trans.}, {\bf 33}, (1985), 223-226.

\bibitem{R178}     
A.G. Ramm,
Calculation of resonances and their
extraction from
 transient fields, {\em J. Math. Phys.}, {\bf 26}, N5, (1985)
 1012-1020.

\bibitem{R190}    
A.G. Ramm,
 {\em Scattering by obstacles}, D.Reidel,
Dordrecht, 1986, pp.1-442.


\bibitem{R196}   
A.G. Ramm, On completeness of the products of
harmonic
 functions, {\em Proc. Amer. Math. Soc.}, {\bf 99}, (1986), 253-256.
 
 
\bibitem{R214} A.G. Ramm, 
Completeness of the products of solutions
to PDE
 and uniqueness theorems in inverse scattering,
 {\em Inverse problems}, {\bf 3}, (1987), L77-L82

\bibitem{R217} A.G. Ramm, 
A uniqueness theorem for two-parameter
inversion,
{\em Inverse Probl.}, {\bf 4}, (1988), L7-10.

\bibitem{R218} A.G. Ramm, 
A uniqueness theorem for a boundary
inverse
 problem, {\em Inverse Probl.}, {\bf 4}, (1988), L1-5.

\bibitem{R220} A.G. Ramm, 
Multidimensional inverse problems and
completeness
 of the products of solutions to PDE, {\em J. Math. Anal.
 Appl.}, {\bf 134}, 1, (1988), 211-253; {\bf 139}, (1989) 302.

\bibitem{R229} A.G. Ramm, 
A simple proof of uniqueness theorem in
impedance
 tomography, {\em Appl. Math. Lett.}, {\bf 1}, N3, (1988),
 287-290.

\bibitem{R233} A.G. Ramm, 
Uniqueness theorems for multidimensional
inverse problems with unbounded coefficients,
{\em J. Math. Anal. Appl.}, {\bf 136}, (1988), 568-574.

\bibitem{R236} A.G. Ramm, 
Multidimensional inverse scattering
problems and
 completeness of the products of solutions to
 homogeneous PDE.,  {\em Zeitschr. f. angew. Math. u.
 Mech.}, {\bf 69}, (1989) N4, T13-T22.

\bibitem{R247} A.G. Ramm, 
Stability of the numerical method for
solving the
 3D inverse scattering problem with fixed energy
 data, {\em Inverse problems}, {\bf 6}, (1990), L7-12.

\bibitem{R248} A.G. Ramm, 
Algorithmically verifiable
characterization of the
 class of scattering amplitudes for small
 potentials,  {\em Appl. Math. Lett.}, {\bf 3}, (1990), 61-65.

\bibitem{R252} A.G. Ramm, 
Completeness of the products of solutions
of PDE
 and inverse problems, {\em Inverse Probl.}, {\bf 6},
 (1990), 643-664.

\bibitem{R254} A.G. Ramm, 
Uniqueness result for inverse problem of
geophysics
 I, {\em Inverse Probl.}, {\bf 6}, (1990), 635-642.

\bibitem{R255} A.G. Ramm, G. Xie 
Uniqueness result for inverse problem of
geophysics II, {\em Appl. Math.
 Lett.}, {\bf 3}, (1990), 103-105.

\bibitem{R257} A.G. Ramm,  J. Sj\"ostrand
An inverse problem for the wave equation,
{\em Math. Zeitschr.}, {\bf 206}, (1991) 119-130. 

\bibitem{R261} A.G. Ramm, Rakesh, 
Property C and an inverse problem for a
hyperbolic
 equation, {\em J. Math. Anal. Appl.},
{\bf 156}, (1991), 209-219. 

\bibitem{R264} A.G. Ramm, 
Stability of the numerical method for
solving 3D
 inverse scattering problem with fixed energy data,
 {\em J.f.die reine und angew. Math}, {\bf 414}, (1991), 1-21.
 
\bibitem{R279} A.G. Ramm, 
Stability of the solution to inverse
scattering
 problem with exact data,
{\em Appl.Math.Lett.}, {\bf 5}, 1, (1992), 91-94
 
\bibitem{R301} A.G. Ramm, 
Approximation by the scattering solutions
and
 applications to inverse scattering,
 {\em Math.Comp.Modelling}, {\bf 18}, N1, (1993), 47-56.
 
\bibitem{R305} A.G. Ramm, 
Scattering amplitude is not a finite-rank
kernel, {\em J. of Inverse and Ill-Posed Problems},
{\bf 1}, N4, (1993), 349-354.(with P.Stefanov)

\bibitem{R306} A.G. Ramm, 
Inverse scattering at fixed energy for
 exponentially decreasing potentials, Proc. of the
 Lapland conference on inverse problems (with
 P.Stefanov). Lecture notes in Phys, N422, Springer-
 Verlag, 1993, 189-192.
 
\bibitem{R309} A.G. Ramm, 
Scattering amplitude is not a finite rank
kernel in
 the basis of spherical harmonics, {\em Appl.Math.Lett.},
{\bf 6}, N5, (1993), 89-92.

\bibitem{R320} A.G. Ramm, 
Inversion of fixed-frequency surface data
for
 layered medium, {\em J. of Inverse and Ill-Posed
 Problems}, {\bf 2}, N3, (1994), 263-268

\bibitem{R330} A.G. Ramm, 
Examples of nonuniqueness for an inverse
problems
 of geophysics, {\em Appl. Math. Lett.}, {\bf 8}, N4, (1995),
87-90.

\bibitem{R228} A.G. Ramm, 
Recovery of the potential from fixed
energy scattering data, {\em Inverse Problems}, {\bf 4}, (1988),
 877-886;  5, (1989) 255.


\bibitem{R241} A.G. Ramm, 
Necessary and sufficient condition for a
scattering
 amplitude to correspond to a spherically symmetric
 scatterer, {\em  Appl.Math.Let.}, {\bf 2}, (1989), 263-265.
 
 
 \bibitem{R249} A.G. Ramm, 
 Is the Born approximation good for solving
the inverse problem when the potential is small?
{\em  J. Math. Anal. Appl.}, {\bf 147}, (1990), 480-485.
 
 
 \bibitem{R260} A.G. Ramm, Symmetry properties for scattering 
amplitudes and
 applications to inverse problems, {\em J. Math. Anal.
 Appl.}, {\bf 156}, (1991), 333-340.
 
 \bibitem{R262} A.G. Ramm, 
 Necessary and sufficient condition for a
PDE to
 have property C, {\em J. Math. Anal.
 Appl.} {\bf 156}, (1991), 505-509.
 
 \bibitem{R278} A.G. Ramm, 
 {\it Multidimensional inverse scattering
problems}, Longman/Wiley, New York, 1992, pp.1-385.
 
 \bibitem{R285} A.G. Ramm, 
 Stability estimates in inverse scattering, {\em Acta
 Appl. Math.},  {\bf 28}, N1, (1992), 1-42.
 
 
 \bibitem{R325} A.G. Ramm, 
 Stability of the solution to inverse
obstacle
 scattering problem, {\em J.Inverse and Ill-Posed
 Problems},  {\bf 2}, N3, (1994), 269-275.
 
 \bibitem{R343} A.G. Ramm, 
 {\em Uniqueness theorems for inverse obstacle
scattering problems in Lipschitz domains, Applic. Analysis}, {\bf 59}, 
(1995),
377-383.
 
 \bibitem{R371} A.G. Ramm, 
 Fundamental solutions to elliptic equations 
with
discontinuous senior coefficients and an inequality
for these solutions, {\em Math. Ineq. and Applic.},  {\bf 1}, N1, (1998), 
99-104.
 
\bibitem{R387} A.G. Ramm,  
Property C for ODE and applications to inverse
scattering, {\em  Zeit. fuer Angew. Analysis}, {\bf 18}, N2, (1999), 
331-348.


\bibitem{R391} A.G. Ramm,  
Inverse problem for an inhomogeneous Schr\"odinger
equation,
{\em Jour. Math. Phys.}, {\bf 40, N8} (1999),3876-3880. 

\bibitem{R392} R. Airapetyan, A.G. Ramm and  A. Smirnova, 
Example of
two different potentials which have practically the same fixed-energy
phase shifts, {\em Phys. Lett A}, {\bf 254, N3-4} (1999), 141-148.  


\bibitem{R393} A.G. Ramm, 
Inverse scattering problem with part of the
fixed-energy phase shifts, {\em Comm. Math. Phys.} {\bf 207, N1} (1999),
231-247.


\bibitem{R402} A.G. Ramm, 
Property C for ODE and applications to inverse
problems, in the book {\em "Operator Theory and Its Applications"}, Amer.
Math. Soc., Fields Institute Communications vol. 25, (2000), pp.15-75,
Providence, RI.

 \bibitem{R421} A.G. Ramm and S.Gutman, 
 Piecewise-constant positive
potentials with practically the same fixed-energy phase shifts, {\em
Applicable Analysis}, {\bf 78, N1-2} (2001), 207-217.



\bibitem{R425} A.G. Ramm,  
Stability of solutions to inverse scattering 
problems with fixed-energy data, {\em Milan Journ. of Math.},
{\bf 70}, (2002), 97-161.

\bibitem{R431} A.G. Ramm, 
Analysis of the Newton-Sabatier scheme for inverting
fixed-energy phase shifts, {\em Applic. Analysis}, {\bf 81, N4} (2002), 965-975.


\bibitem{R433} A.G. Ramm,  
A counterexample to a uniqueness result,
{\em Applic. Analysis}, {\bf 81, N4} (2002), 833-836.

\bibitem{R445} A.G. Ramm,  
Comments on the letter of P.Sabatier,
http://arXiv.org/abs/math-ph/0308025,  PaperId: math-ph/0308025.


\bibitem{R460} A.G. Ramm, 
One-dimensional inverse scattering and spectral
problems, {\em Cubo a Mathem. Journ.}, {\bf 6, N1} (2004), 313-426.

 
\bibitem{R470} A.G. Ramm, 
{\it Inverse problems}, Springer, 
New York, 2005.

\bibitem{R476} A.G. Ramm, 
{\it Wave scattering by small bodies of
arbitrary shapes}, World Sci. Publishers, Singapore, 2005.


\bibitem{R515} A.G. Ramm, 
Distribution of particles which produces a
"smart" material, {\em Jour. Stat. Phys.}, {\bf 127, N5} (2007), 915-934. 

\bibitem{R517} A.G. Ramm, 
Inverse scattering problem with data at fixed 
energy and fixed incident direction, Nonlinear Analysis: Theory, Methods 
and Applications,  doi:10.1016/j.na.2007.06.047

\bibitem{R518} A.G. Ramm,  
Many-body wave scattering by small bodies, 
{\em J. Math. Phys.},
{\bf 48, N2} 023512, (2007). 

\bibitem{R524} A.G. Ramm, 
Wave scattering by small particles in a medium,
{\em Phys. Lett.} {\bf A367}, (2007), 156-161.

\bibitem{R525} A.G. Ramm, 
Wave scattering by small impedance particles 
in a medium, {\em Phys. Lett.} {\bf A368, N1-2} (2007), 164-172.

\bibitem{R528} A.G. Ramm, 
Scattering by many small bodies and applications
to condensed matter physics, {\em EPL} (Europ. Physics Lett.) (to appear).

\bibitem{R529} A.G. Ramm,  
Many-body wave scattering by small bodies and 
applications, {\em J. Math. Phys.}, {\bf 48}, 10, (2007)

\bibitem{R531} A.G. Ramm,  
Fixed-energy inverse scattering,
Felicitation volume for Prof. V. Lakshmikantham, Cambridge Univ. 
Press,
2008, (to appear).


 





\end{thebibliography}
\end{document}